\documentclass[conference, a4paper]{IEEEtran}

\usepackage{amsmath,graphicx, amssymb, amsfonts}
\usepackage{booktabs}
\usepackage[bookmarks=false]{hyperref}
\usepackage[T1]{fontenc}
\usepackage[font=footnotesize]{caption}
\usepackage{color,soul}
\usepackage{float}
\usepackage{url}

\definecolor{gray}{rgb}{1,0.6,0.5}
\setstcolor{gray}
\sethlcolor{gray}
\usepackage{multirow}
\usepackage{siunitx}
\usepackage{enumitem}
\usepackage{cite}
\usepackage{subcaption}
\usepackage{dsfont}
\usepackage{diagbox}
\usepackage{array}
\usepackage[thinc]{esdiff}
 \setlist[itemize]{leftmargin=*}

\begin{document}
\title{VMAF-based Bitrate Ladder Estimation for Adaptive Streaming}

\author{Angeliki~V.~Katsenou$^*$, Fan Zhang$^*$, Kyle Swanson$^\dagger$, Mariana Afonso$^\dagger$, Joel Sole$^\dagger$,  and David~R.~Bull$^*$\\
$^*$Visual Information Lab, University of Bristol, Bristol, BS1 5DD, UK\\
%\{angeliki.katsenou, fan.zhang, dave.bull\}@bristol.ac.uk\\
$^\dagger$Netflix, Inc, Los Gatos, California, USA\\
%\{jsole,kswanson,mafonso\}@netflix.com\\
}


\maketitle

\begin{abstract}
 In HTTP Adaptive Streaming, video content is conventionally encoded by adapting  its spatial resolution and quantization level to best match the prevailing network state and display characteristics. It is well known that the traditional solution, of using a fixed bitrate ladder, does not result in the highest quality of experience for the user. Hence, in this paper, we consider a content-driven approach for estimating the bitrate ladder, based on spatio-temporal features extracted from the uncompressed content. The method implements a content-driven interpolation. It uses the extracted features to train a machine learning model to infer the curvature points of the Rate-VMAF curves in order to guide a set of initial encodings. We employ the VMAF quality metric as a means of perceptually conditioning the estimation. When compared to exhaustive encoding that produces the reference ladder, the estimated ladder is composed by 74.3\% of identical Rate-VMAF points with the reference ladder. The proposed method offers a significant reduction of the number of encodes required, 77.4\%, at a small average Bj{\o}ntegaard Delta Rate cost, 1.12\%. 
\end{abstract}


% \fi
\IEEEpeerreviewmaketitle

\section{Introduction}
\label{sec: Introduction}
The importance of visual communications in our daily activities and interactions has increased dramatically in recent years, not least due to restrictions imposed by the global COVID-19 pandemic. We are all creating and consuming increased volumes of video data with video streaming companies reporting major increases in video downloads shortly after WHO declared COVID-19 as a pandemic~\cite{BitmovinReport}. 

HTTP Adaptive Streaming (HAS) is a process employed by most video services to address  dynamically changing network conditions. In Dynamic Adaptive Streaming over HTTP (DASH)~\cite{MPEG-Dash}, video content is encoded by varying spatial resolution and quantization level in order to adapt to the changing state of a heterogeneous network and to differing display device specifications. For example, if a streaming client monitors a change in the rate of an incoming video chunk that cannot support a smooth (without re-buffering) play-out, it will signal the need to switch to a stream at a lower bitrate. To this end, the creation of a set of video encodings at different bitrates is required at the server. This set of encodings are normally represented using a bitrate ladder. The traditional HAS solution uses a fixed bitrate ladder (a set of fixed bitrate-resolution pairs) but this approach cannot ensure a high quality of experience for all types of video content. 

An improvement over this fixed solution is to introduce differentiation based on content genre, e.g.~\cite{LedererMMSys2012}. For example, higher bitrates can be used for sports content with rapid motion and fast scene changes. Previous solutions, however, were not tailored to video content characteristics, resulting in noticeable visual artifacts.

Recently, content-customised solutions have been reported and adopted by industry, such as those used by Netflix~\cite{AaronNtflx2015,NetflixDOICIP2016, DynOptimiser, NtflxBlog2020}. The key task is to invest in pre-processing where each video title is split into shorter clips or chunks, usually associated with shots. Each short video chunk is encoded using optimized parameters, i.e. resolution, quantization level, intra-period, etc., with the aim of building the Pareto Front (PF) across all Rate-Quality curves. Then a set of target bitrates is used to find the best encoded bitstreams. The quality metric used for this in the Netflix case is Video Multi-method Assessment Fusion (VMAF)~\cite{VMAFblog}. Given the extensive parameter space (compression levels, spatial and temporal resolution, codec type etc.) and taking into account the fact that this process must be repeated for each video chunk, the amount of computation needed is massive. As a consequence, the industry heavily relies on cloud computing services, and this comes at a high cost in financial, time and compute terms.

Many other approaches that provide content-driven customisation have been proposed recently.
Most of these methods first conduct a complexity analysis. An approach reported by Bitmovin~\cite{BitmovinBitrateLadder,TimmererICME2020}, performs a complexity analysis on each incoming video and inputs that into a machine-learning model to adjust the encoding profile to match the content. CAMBRIA~\cite{Cambria}, estimates the encoding complexity by running a fast constant rate encoding~\cite{Cambria}. In~\cite{Giladi2018}, trial encodes are used to collect coding statistics at low resolutions and these are utilized within a probabilistic framework to improve encoding decisions at higher resolutions.
MUX~\cite{MUX} introduced a deep-learning based approach that takes, as input, the vectorized video frames and predicts the bitrate ladder.
Another interesting approach that takes into account both quality constraints and bitrate network statistics was proposed by Brightcove~\cite{Reznik2018,ReznikICME2019}. The quality metric used in this case was the Structural Similarity Index Measure (SSIM) and bitrate constraints were based on probabilistic models. Finally, recently iSize~\cite{iSize} proposed the use of pre-encodes within a deep learning framework to decide on the optimal set of encoding parameters and resolution at a block level. 

While all of the above solutions are significant and have contributed in the enhancement of video services, it is not possible to make direct detailed comparisons as they are proprietary. In our previous work~\cite{KatsenouPCS2019}, we predicted the intersection points of the PSNR-Rate curves. Then, in~\cite{Katsenou2021}, we extended the method to the estimation of the bitrate ladder by using encodings at the intersection points to estimate the Pareto Front (PF) parameters, resolution and quantization parameters, at the target bit rates.  

In this paper, we propose a new content-driven method that offers an improved bitrate ladder estimation based on VMAF. VMAF has been shown to exhibit a better correlation with perceptual quality than PSNR; hence the resulting bitrate ladders should deliver perceptually improved video streams. 
The method makes a feature-based prediction of the highest curvature points of the Rate-VMAF curves to guide a small set of initial encodings close to the area of interest. The results show significant improvement in terms of the number of required computations for only a small mean Bj{\o}ntegaard Delta Rate (BD-Rate)~\cite{Bjontegaard} cost.

The remainder of this paper is structured as follows. Section~\ref{sec: CurvesAndChar} describes the dataset and the Rate-VMAF curve characteristics. In Section~\ref{sec: RefLadder}, the the definition of the reference bitrate ladder is provided. The proposed framework and the evaluation results are elaborated in Sections~\ref{sec: ProposedMethod}-\ref{sec: Evaluation}. Finally, conclusions are summarised in Section~\ref{sec: Conclusion}.

\section{Rate-VMAF Curves and Characteristics}
\label{sec: CurvesAndChar}

\subsection{Description of the Dataset}
\label{ssec: dataset}
We employed the same dataset of 100 publicly available UHD video sequences as in our previous work~\cite{KatsenouPCS2019,AfonsoSPIE2018, DODataset}. The sequences have a native resolution of 3840$\times$2160, the chroma format is 4:2:0, the bit depth is 10, and the frame rate 60 fps. Each sequence contains a single scene (no scene cuts) including a variety of different objects/scenes/regions of interest, camera motions, colours, and spatial activity. In this paper, we consider \{2160p, 1080p, 720p, 540p\} as the set of test resolutions used to develop  and validate our methods. We use the Lanczos-3 filter~\cite{Duchon} for spatial down/up-sampling throughout.

\subsection{Rate-VMAF Curves across Resolutions and the Pareto Front}

\begin{figure}[!t]
\begin{minipage}[b]{.49\linewidth}
\centering
\includegraphics[width=\linewidth]{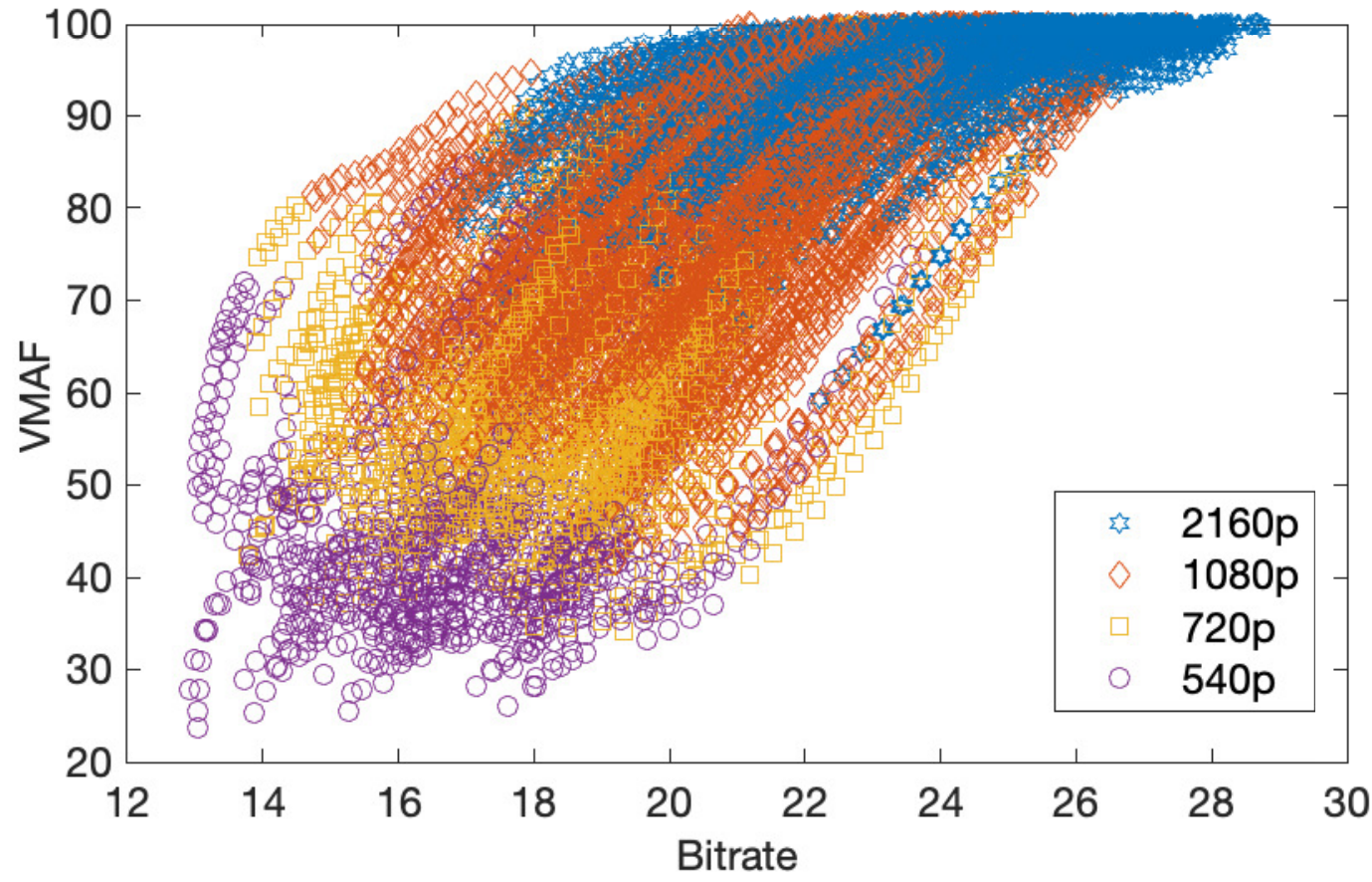}
\footnotesize{(a) $\log$(R)-VMAF PFs.}
%\vspace{.1cm}
\end{minipage}
\hfill
\begin{minipage}[b]{.49\linewidth}
\centering
\includegraphics[width=\linewidth]{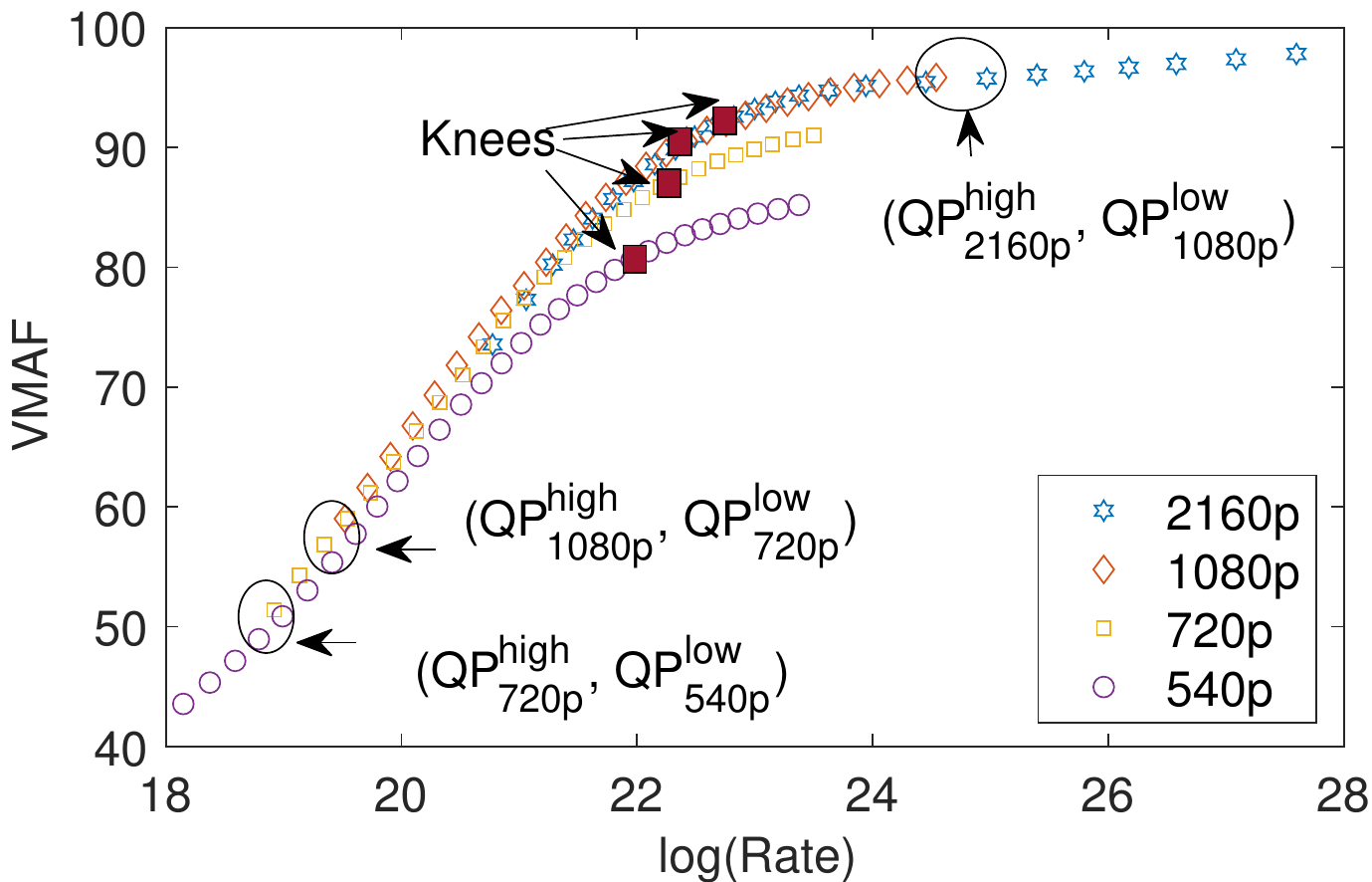}
\footnotesize{(b) Cross-over QP and knee points.}
%\vspace{.1cm}
\end{minipage}
\caption{Rate-VMAF curves and Pareto fronts of the considered dataset across four spatial resolutions: 2160p, 1080p, 720p, and 540p.}
\vspace{-.5cm}
\label{fig:CH_VMAF}
\end{figure}

Rate-VMAF curves exhibit characteristics that are different to those produced by other quality metrics. For consistency with previous work and ease of visualization, we convert these first to the $\log$(Rate) domain.
In Fig.~\ref{fig:CH_VMAF}, we illustrate the resulting Pareto surfaces for our dataset across the four spatial resolutions. It is clear that working in the $\log$(Rate) domain is beneficial as the curves become smoother. Besides this, the saturation of VMAF at high bitrates and high resolutions is evident. This characteristic is content dependent and can be exploited when building a bitrate ladder.

\subsection{Intersection Points of Rate-VMAF Curves}
An important characteristic of the Pareto Front (PF), used for constructing the bitrate ladder, is the set of points where resolution switches, i.e. the intersection points of the Rate-VMAF curves. It is more practical to define these intersection points as pairs of QP values, called cross-over QPs~\cite{Katsenou2021}, 
   $\left(QP^{high}_{s},QP^{low}_{s-1}\right)$, 
with $s \in \mathcal{S}$ resolutions of the intersecting curves of the same video sequence. $level \in \{\textrm{high}, \textrm{low}\}$ defines the range of QPs. The resolution and level cannot be the same for both QPs in a pair. For example, in Fig.~\ref{fig:CH_VMAF}, the pair ($QP^{high}_{1080p},QP^{low}_{720}$) represents the 1080p intersection with the 720p curve.

 \subsection{Knee Points of Rate-VMAF Curves}
 An important characteristic of a Rate-VMAF curve is the point of highest curvature or ``knee'' point, $K$. This gives an indication of when the rate of improvement of the video quality will start decreasing, as shown in Fig.~\ref{fig:CH_VMAF}(b). We use the Kneedle algorithm, as described in~\cite{Kneedle}, to compute the knee points of the curves across the resolutions. This algorithm is based on the notion that the points of maximum curvature in a dataset are approximately the set of points in a curve that are local maxima, if the curve is rotated clockwise by an angle defined by the line that connects the lowest and highest values in the dataset.
 
\begin{figure}[!t]
 \begin{minipage}{.49\linewidth}
\centering    
\includegraphics[width=\linewidth]{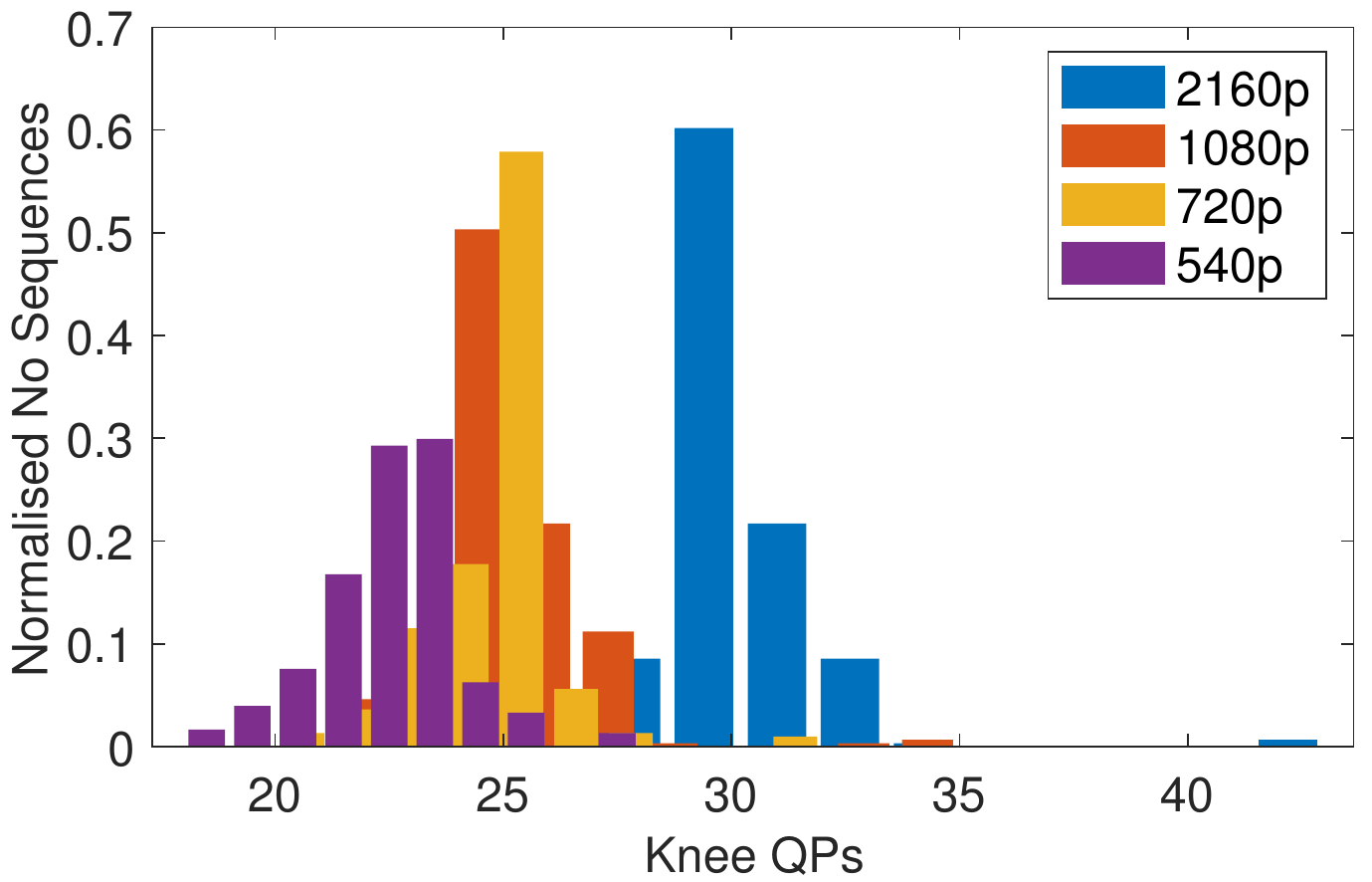}
 \footnotesize{(a) Distribution of knee QPs.}
\vspace{.1cm}
\end{minipage}
\hfill
 \begin{minipage}{.49\linewidth}
\centering    
\includegraphics[width=\linewidth]{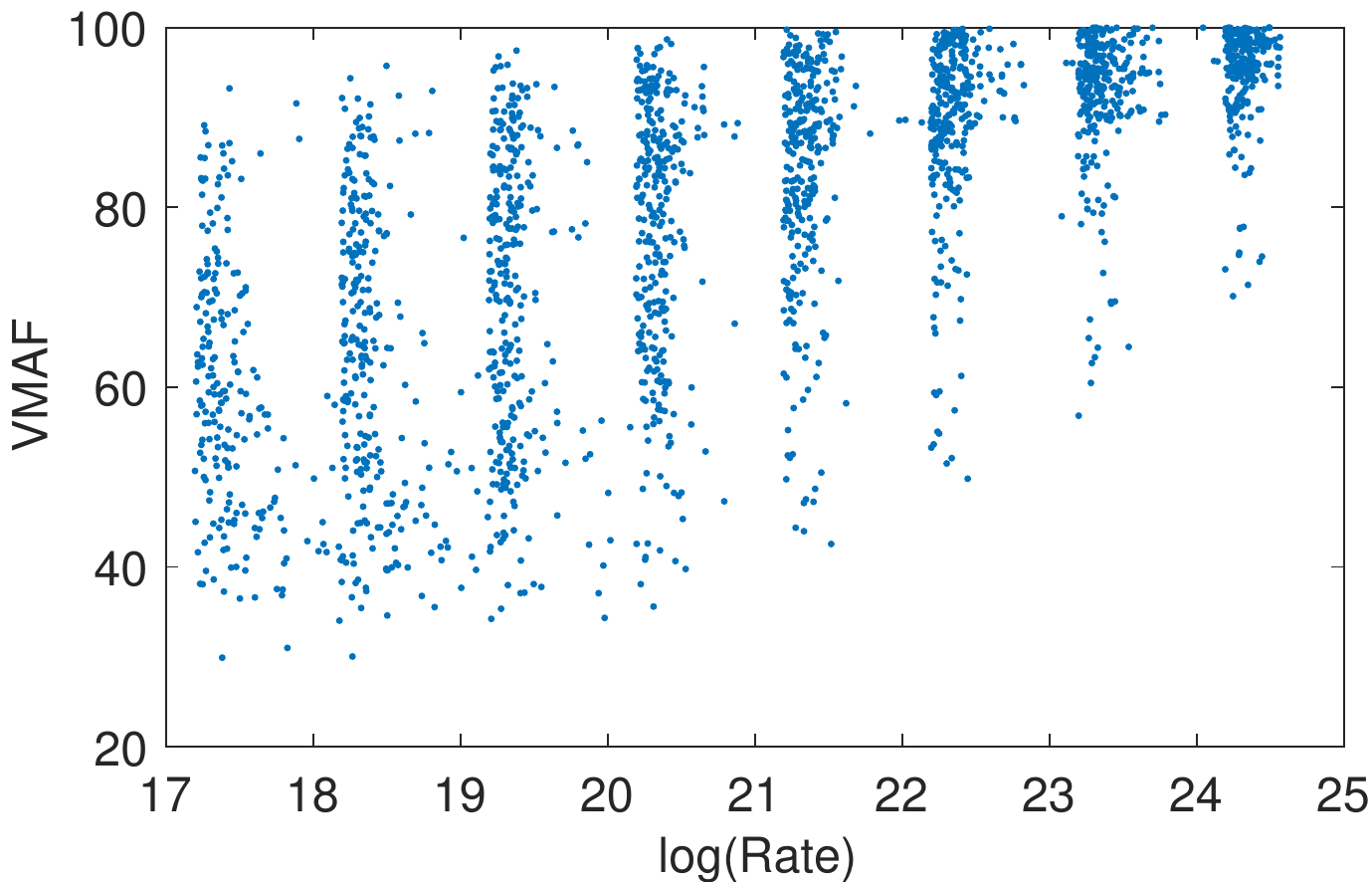}
 \footnotesize{(b) Ladder points for all dataset.}
\vspace{.1cm}
 \end{minipage}
\caption{Knee points and reference VMAF bitrate ladder for all dataset.}
\vspace{-0.4cm}
\label{fig:BirateLadder}
\end{figure}

 As shown in Fig.~\ref{fig:BirateLadder}, 
 the distributions of the knee QPs for the different sequences are quite tight around their mean values: 30.00$\pm$1.60 for 2160p, 24.99$\pm$1.72 for 1080p, 24.87$\pm$1.50 for 720p., and 23.08$\pm$1.52 for 540p. It is also important to note that although the knee points of the higher resolutions are usually part of the PF, the knee points of the lower resolutions are typically not part of it. This can be observed in the example of  Fig.~\ref{fig:CH_VMAF} (b).

\section{The VMAF Bitrate Ladder}
\label{sec: RefLadder}
We first perform exhaustive encodings across resolutions for a wide range of QP values to construct the optimal $\log$(Rate)-VMAF Pareto Front (which will serve as our reference) and then determine the intersection points of the $\log$(Rate)-VMAF curves between different spatial resolutions. These intersection points mark the limits of the bitrate range for which encoding at the given resolution yields the best quality~\footnote{When encoding at a lower resolution, all metrics are computed on the upscaled version: all sequences are first downscaled, then encoded, decoded and, finally, upscaled to the native resolution prior to metric computation.}.

\subsection{Definition}
The initial step in constructing the bitrate ladder selects the target bitrates that will represent the rungs, i.e. $\mathcal{R_L}=\{R_{L,1},R_{L,2},\ldots,R_{L,|\mathcal{L}|}\}$, where $|\mathcal{L}|$ is the cardinality of $\mathcal{R_L}$ and $R_{L,1}<R_{L,2}<\ldots<R_{L,|\mathcal{L}|}$. The VMAF bitrate ladder is fully defined as a set of tuples $\mathcal{L}$ that comprise bitrate values $\mathcal{R_L}$, the associated set of VMAF values $\mathcal{V_L}$, a set of QP values $\mathcal{QP_L}$, and a set of resolutions $\mathcal{S_L}$, i.e.
\begin{equation}
\label{eq: Ladder}
   \mathcal{L}:= \{\langle R_{L,i},V_{L,i},QP_{L,i},S_{L,i} \rangle \}^{|\mathcal{L}|}_i \; , 
\end{equation}
\noindent
with
 $R_{L,1}<R_{L,2}<\ldots<R_{L,|\mathcal{L}|}$, $V_{L,1}\leq V_{L,2}\leq \ldots \leq V_{L,|\mathcal{L}|}$, $S_{L,1}\leq S_{L,2}\leq \ldots \leq S_{L,|\mathcal{L}|}$. 
The above constraints ensure the monotonicity of the set of parameters $\mathcal{R_L}$, $\mathcal{V_L}$, and $\mathcal{S_L}$. For the set $\mathcal{QP_L}$, monotonicity is assumed for $QP_{L,i}$ values of the same resolution.

\subsection{Building the Reference Ladder}
\label{ssec: RefLadder}
In order to construct the bitrate ladder, we sample the Pareto front using the set of target bitrates $\mathcal{R_L}$. From the resulting points, we check whether it is meaningful to retain all ladder rungs if we cannot significantly improve quality. Therefore, we monitor the slope of the sampled Rate-VMAF points so that:   
%\begin{equation}
    $\diff{V_{\textrm{L,}i}}{R_{\textrm{L}}} > \epsilon \;$
%\end{equation}
when $V_{\textrm{L,}i}>V_{high}$, where $\epsilon\in \mathds{R}, \epsilon \rightarrow 0$ and $V_{high}=97$. 
As a consequence of the above constraint, the length of the ladder might vary. The use of variable ladder lengths was suggested before in~\cite{Reznik2018} and is dependent on content features and their relation to perceptual quality.

We considered the [150kbps,25Mbps] bitrate range for the ladder and that each new bitrate rung is twice that of the previous one, i.e. $R_{L,i}=2 R_{L,i-1}$. As can be seen in Fig.~\ref{fig:BirateLadder}, the eight rungs on the ladder are clearly visible and are shifted to a greater or lesser extent according to the sequence. As VMAF is bounded by a maximum value of 100, the points become increasingly dense at higher VMAF values and higher bitrates. For about half (49\%) of the tested sequences, there are fewer than eight rungs on the ladder  -  typically associated with sequences that can reach to a visual quality equivalent to the original (according to VMAF). These could be static sequences, with low amounts of structural or textural information that require lower bitrates for high quality reconstruction.

\section{Proposed Method}
\label{sec: ProposedMethod}
Previous work~\cite{KatsenouPCS2019, Katsenou2021} showed that the best performing method in terms of BD-Rate cost was the interpolation-base method. 
The method proposed in this paper is based on encoding using only a subset of QP values per resolution. The selection of the subset is content-driven and is related to the knee point of the curve. After encoding, piece-wise cubic Hermite interpolation~\cite{pchip} is applied to for the interim QPs. Based on  these values, the PF is extracted. This method produces a suboptimal  solution,  whose  accuracy  depends  on  the number of encodes performed per resolution. The added benefit of this method is that it significantly reduces the number of encodings required compared to exhaustively encoding at all QPs.

\begin{figure}[!t]
\begin{minipage}[b]{.49\linewidth}
\centering
\includegraphics[width=\linewidth]{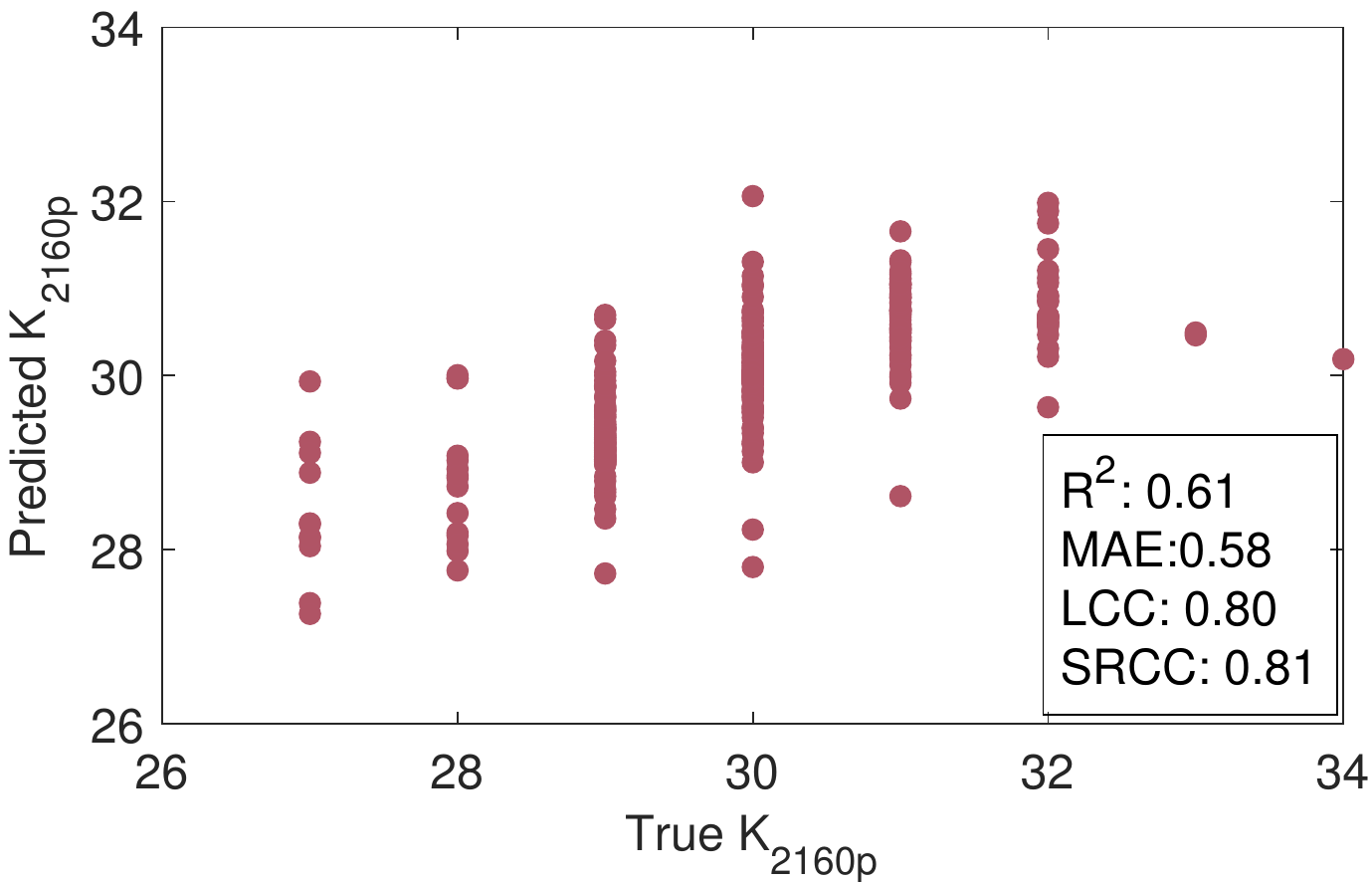}
\footnotesize{(a) $\widehat{K}_{2160p}$.}
%\vspace{.1cm}
\end{minipage}
\hfill
\begin{minipage}[b]{.49\linewidth}
\centering
\includegraphics[width=\linewidth]{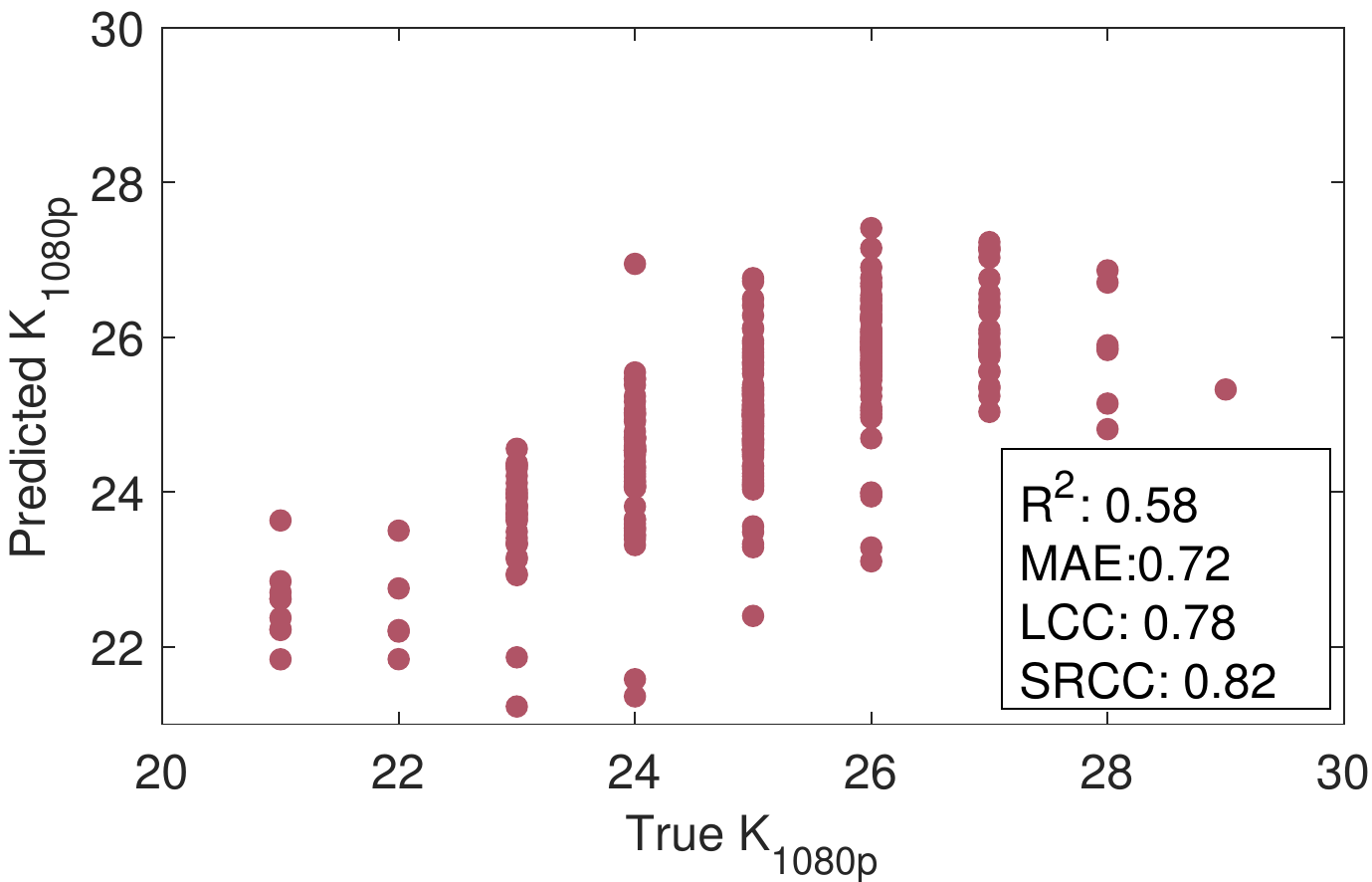}
\footnotesize{(b) $\widehat{K}_{1080p}$.}
%\vspace{.1cm}
\end{minipage}
\begin{minipage}[b]{.49\linewidth}
\centering
\includegraphics[width=\linewidth]{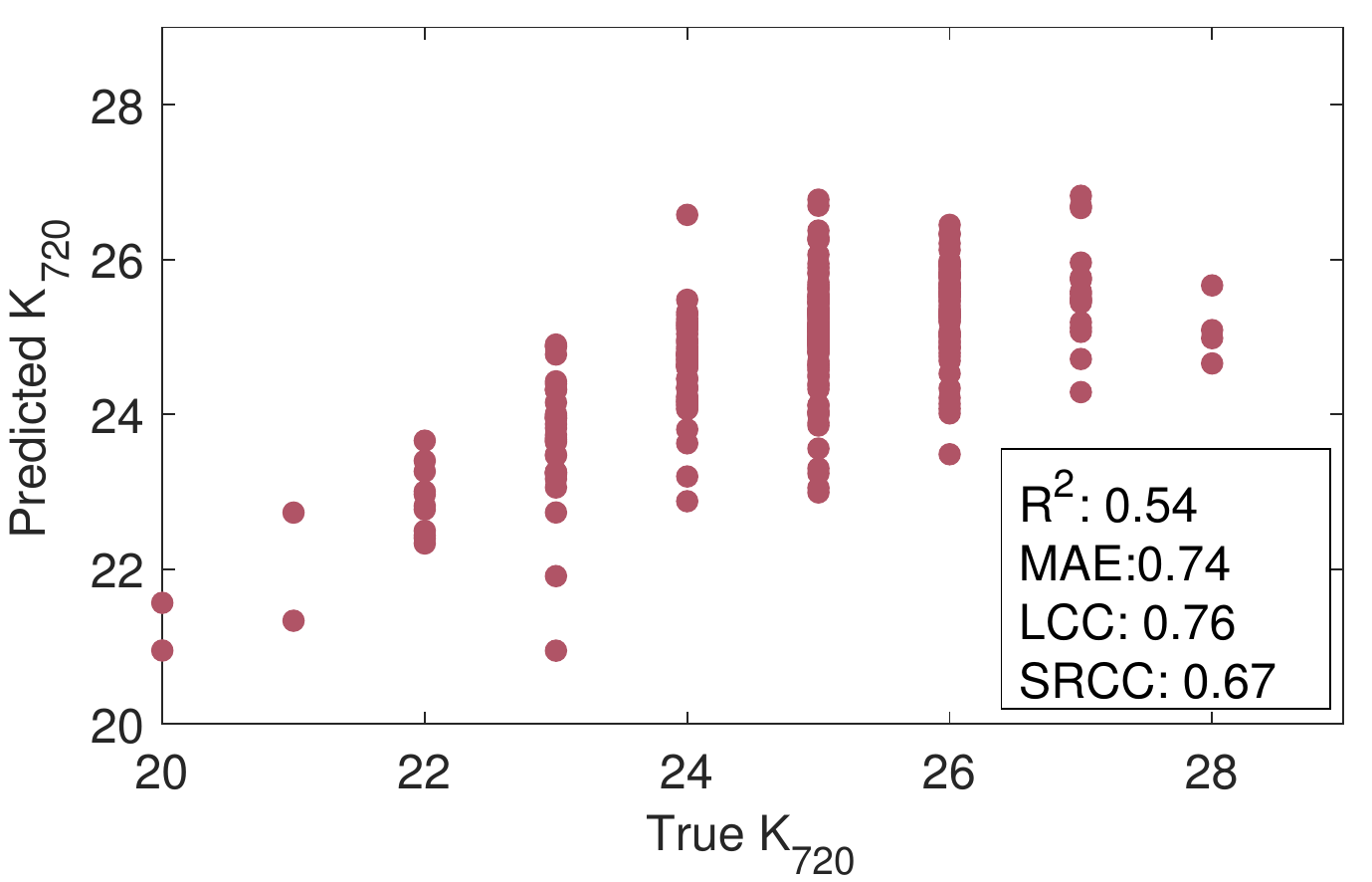}
\footnotesize{(c) $\widehat{K}_{720p}$.}
%\vspace{.1cm}
\end{minipage}
\hfill
\begin{minipage}[b]{.49\linewidth}
\centering
\includegraphics[width=\linewidth]{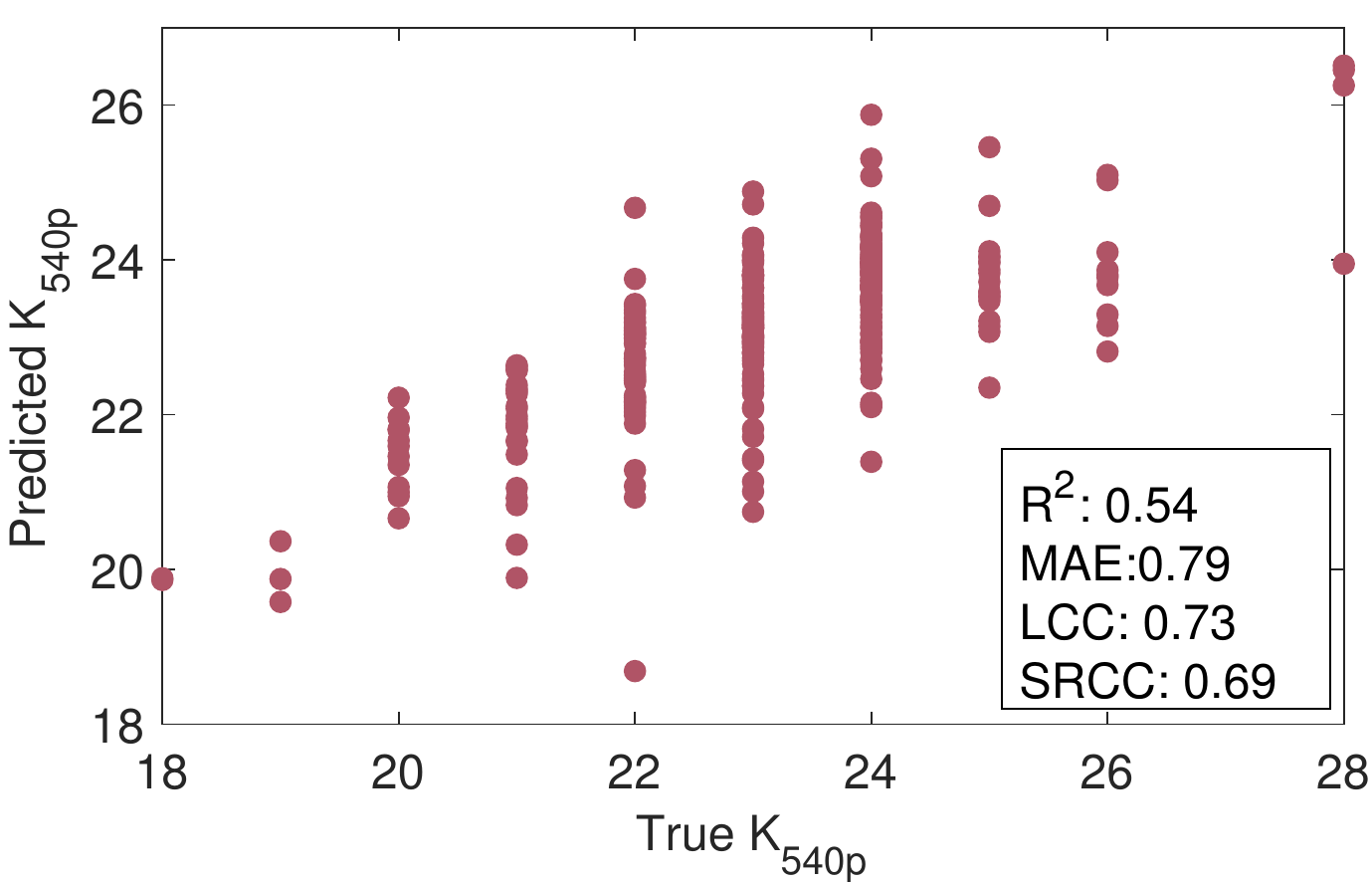}
\footnotesize{(d) $\widehat{K}_{540p}$.}
%\vspace{.1cm}
\end{minipage}
\caption{Predicted Knee Points against the ground truth. Selected features used for prediction: F1-2, F4-5, F7, F9-10, F12, F15-17.}
\vspace{-0.5cm}
\label{fig: PredKnees}
\end{figure}

In this work, we propose a Content-driven Interpolation-based Ladder (CIL) estimation method. This method uses content features to estimate the knee of the curve at each resolution for each sequence. 
Spatio-temporal features are extracted first to predict the knee QPs.
We followed a sequential prediction of the knee QPs starting from the highest resolution down to the lowest. At each step, we applied feature selection, and particularly recursive feature~\cite{Kuhn2013}. Next we trained and tested several machine-learning regression methods, including Support Vector Machines with different kernels and Random Forests, finding that Gaussian Processes (GP), with a 5/2 Mat\'{e}rn covariance kernel~\cite{Rasmussen2005}, performed best for this work. To avoid overfitting, we deployed a ten-fold random cross-validation process. The results from the ten-fold cross-validation are shown in Fig.~\ref{fig: PredKnees}. Despite the fact that R$^2$, LCC, and SRCC values are not as very high, the MAE is small (<0.79), which is adequate to yield good results for the bitrate ladder estimation, as shown in Section~\ref{sec: Evaluation}. 

Next, the knees are used to subsample the QP range that falls within the PF. As observed for the higher resolutions, the knee points have a high probability of belonging to the PF, while this is less probable for the lower resolutions, we determine the initial encodes by evenly spacing a number $n \in \mathds{N}$ of QPs within the following ranges:
\begin{equation}
      QP_{j,s}\in [K_{j,s}+t_s,QP_{\max}],   
\end{equation}
where $j$ denotes the sequence, $t_s \in \mathds{Z}$ is an offset, and $\quad s \in \{2160p,1080p,720p,540p\}$.

After encoding on the QPs above, piece-wise cubic Hermite interpolation is applied to find the Rate-VMAF values for the interim QPs. The bitrate ladder is constructed based on the encodings using the estimated QPs at the target bitrates.

\section{Predicting the Bitrate Ladder}
\label{sec: Evaluation}
The HEVC reference software (version HM 16.20) was employed in this study using its Random Access mode, a 64-frame Intra Period and a Group of Pictures (GoP) length of 16 frames~\cite{HEVCpaper,CTC2017}. After encoding, decoding, and upscaling the spatial resolution to 2160p, we computed VMAF and bitrate at a GoP level that enabled a larger coverage of the Rate-VMAF space.

\subsection{Content Features}
From the vast variety of low-level content features, we employ those spatio-temporal features that have been successfully used in our previous related work~\cite{Katsenou2021, KatsenouPCS2019}:
\begin{itemize}
    \item Gray Level Co-occurrence Matrix (GLCM)~\cite{Haralick1973} mean descriptors across frames: contrast F1.$\textrm{meanGLCM}_{\textrm{con}}$; correlation F2.$\textrm{meanGLCM}_{\textrm{cor}}$; homogeneity F3.$\textrm{meanGLCM}_{\textrm{hom}}$; energy F4.$\textrm{meanGLCM}_{\textrm{enr}}$; entropy F5.$\textrm{meanGLCM}_{\textrm{ent}}$.
    \item Temporal Coherence (TC)~\cite{KatsenouPCS2016} with its interframe statistics: mean F6.$\textrm{meanTC}_{\textrm{mean}}$; standard deviation F7.$\textrm{meanTC}_{\textrm{std}}$; skewness F8.$\textrm{meanTC}_{\textrm{skw}}$; kurtosis F9.$\textrm{meanTC}_{\textrm{kur}}$; and entropy F10.$\textrm{meanTC}_{\textrm{entr}}$, all expressed as a mean across frames.
    \item Interframe Normalised Cross-Correlation (NCC)~\cite{KatsenouMMSP2017} statistics: mean F11.$\textrm{meanNCC}_{\textrm{mean}}$; standard deviation F12.$\textrm{meanNCC}_{\textrm{std}}$; skewness F13.$\textrm{meanNCC}_{\textrm{skw}}$; kurtosis F14.$\textrm{meanNCC}_{\textrm{kur}}$; and entropy F15.$\textrm{meanNCTC}_{\textrm{entr}}$, all expressed as a mean across frames. 
    \item Mean Squared Error of the spatial Rescaling (RsMSE) of the first frame~\cite{AfonsoSPIE2018}: F16.$\textrm{RsMSE}_\textrm{1080p}$ (from 2160p to 1080p), F17.$\textrm{RsMSE}_\textrm{720p}$ (from 2160p to 720p).
\end{itemize}

\subsection{Compared Methods}
Because there are no publicly available implementations of the proprietary solutions described in Section~\ref{sec: Introduction}, we have considered and tested the methods described below. 
\subsubsection{Reference Ladder (RL)} This exhaustive search approach was used to construct our reference bitrate ladder as explained earlier. All sequences were encoded at QP values within $\{15, 16, \ldots, 45\}$ range. The reference bitrate ladder was constructed as explained in Section~\ref{ssec: RefLadder}.

\subsubsection{Naive Interpolation-based Ladder (NIL)} This method is based on encoding using only seven QP values per resolution, as in~\cite{Katsenou2021}. After encoding, a piece-wise cubic Hermite interpolation~\cite{pchip} is used to estimate the Rate-VMAF values for the interim QPs. Based on these estimated points, the ladder is constructed by encoding at the closest QP to the target bitrates.

\subsubsection{Content-driven Interpolation-based Ladder (CIL)} 
 This is the proposed method as described above. To determine the offset $t_s $ values, the ranges of the distributions of cross-over QPs were combined with the distributions of the knee points. The presented results are for $t_\textrm{2160p}=t_\textrm{1080p}=-4$, $t_\textrm{720p}=6$, and $t_\textrm{540p}=10$. 
Furthermore, we explored the impact of the different number of initial encodes required per resolution $n \in \{4,5,6,7\}$ and named accordingly the versions, CIL-$n$.
We only considered up to 7 encodes in order to directly compare to NIL that uses seven initial encodes per resolution.

\subsubsection{Feature-based Predicted Ladder (FL)} 
The same spatio-temporal features as in CIL case are used. Similarly, feature selection, training and sequential prediction of the cross-over points per resolution takes place. GPs are also employed here with a ten-fold cross-validation.
Then encodings at the cross-over QPs and on additional points are used to define linear models that help estimating the QP at the target bitrate. Further to the encodes at the six cross-over QPs, we require two additional in order to determine the unknown parameters of the linear model at each resolution. Particularly, we require one more encode for the 2160p and one for the 540p. The QP value selection for the extra encodes for each sequence $j$ is decided as below:
\begin{equation}
     QP_{s}=
\begin{cases}
    \widehat{QP}^{level}_{j,s}-\delta & \text{if } \widehat{QP}^{level}_{j,s}\geq QP_{m}\\
    \widehat{QP}^{level}_{j,s}+\delta,              & \text{otherwise}
\end{cases}
\end{equation}
where $\delta, \; QP_m\in\mathds{N}$, $s \in$ \{2160p, 540p\}, and $level \in \{low, high\}$. The $\delta$s have been selected based on the distributions of ground truth cross-over QPs. In the presented results, $QP_{m}=30$, $\delta=5$ for 2160p and $QP_{m}=38$, $\delta=2$ for 540p.

\subsection{Evaluation of the Predicted Ladders}
\label{ssec: Results}
We tested on a bitrate range typical for video streaming for the considered resolutions, from 150kbps to 25Mbps.
We evaluated the proposed methods by computing the BD metrics (mean and mean absolute deviation (mad)\footnote{We selected the mean absolute deviation (mad) instead of standard deviation because the BD-Rate distributions are not normal.} of BD-Rate) against RL, the maximum number of encodes required per method, and the percentage of the estimated ladder points that are identical to RL points, RL-hits. 

\begin{figure}[!t]
\begin{minipage}{.49\linewidth}
\centering    
\includegraphics[width=\linewidth]{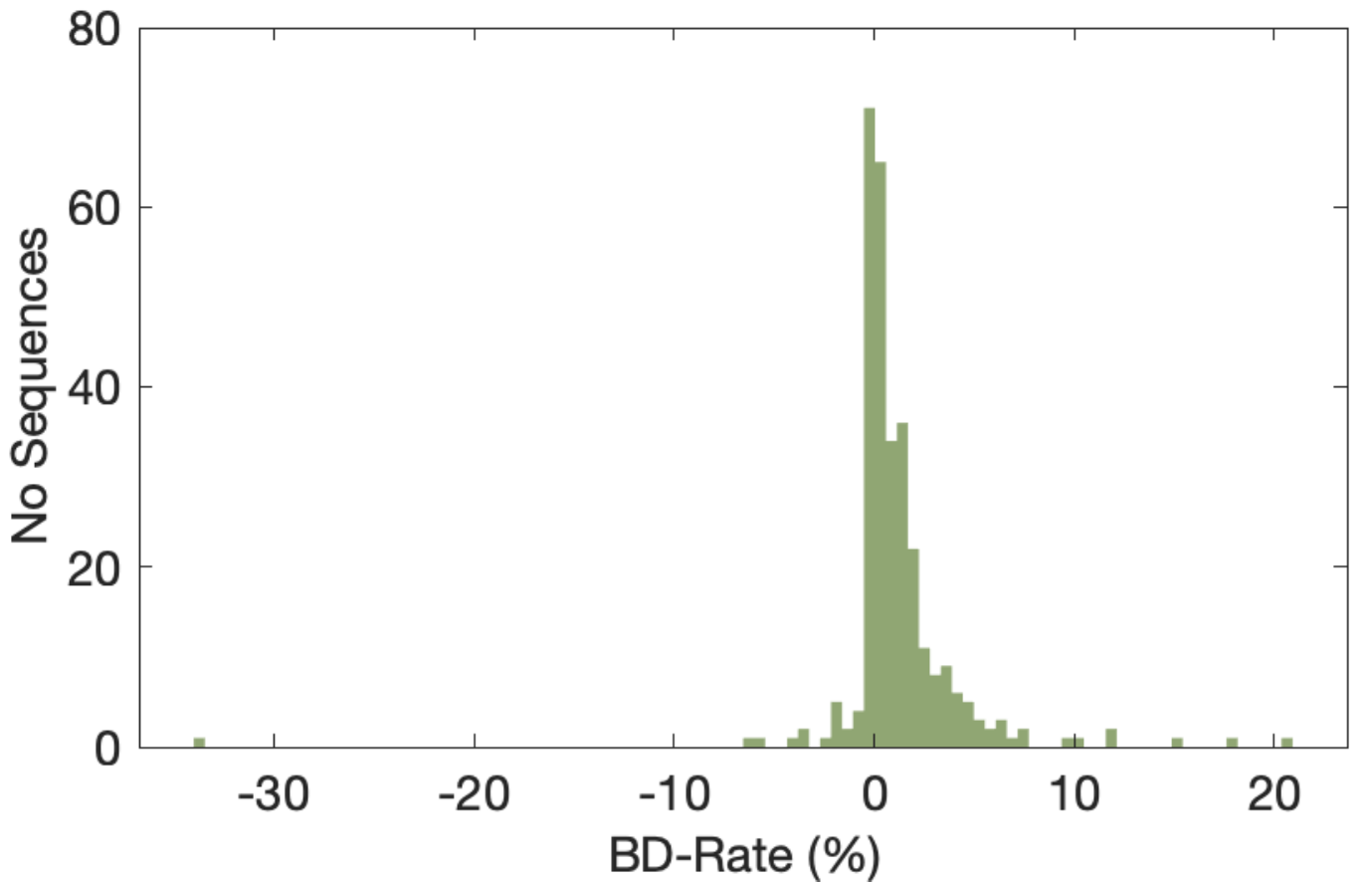}
\footnotesize{(a) NIL vs RL method - BD-Rate: mean 1.1955\%, mad 1.6892\%.}
\vspace{.1cm}
 \end{minipage}
 \hfill
  \begin{minipage}{.49\linewidth}
\centering    
\includegraphics[width=\linewidth]{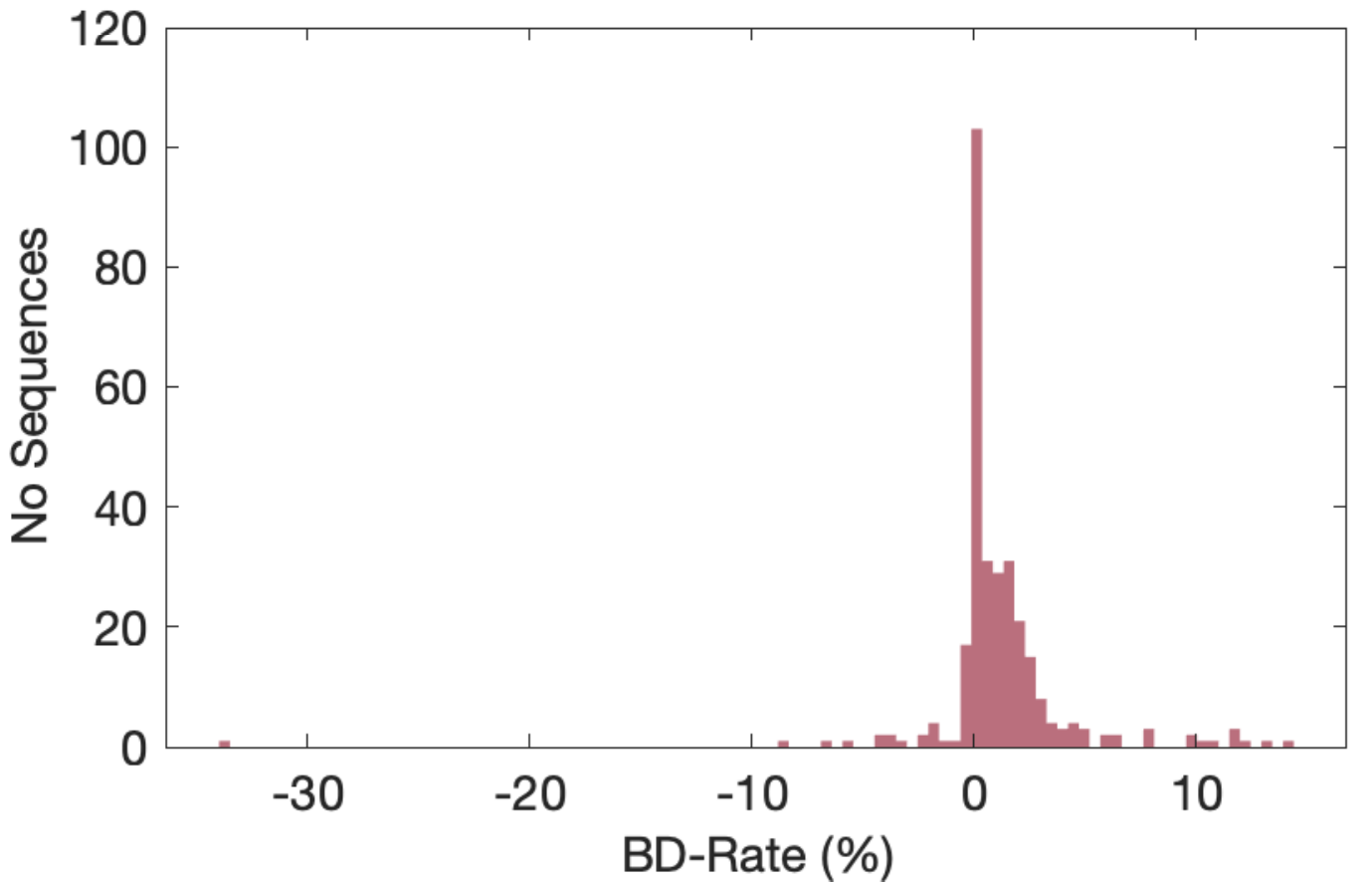}
\footnotesize{(b) CIL-5 vs RL method - BD-Rate: mean 1.1214\%, mad 1.7017\%.}
\vspace{.1cm}
 \end{minipage}
 
 \begin{minipage}{.49\linewidth}
\centering    
\includegraphics[width=\linewidth]{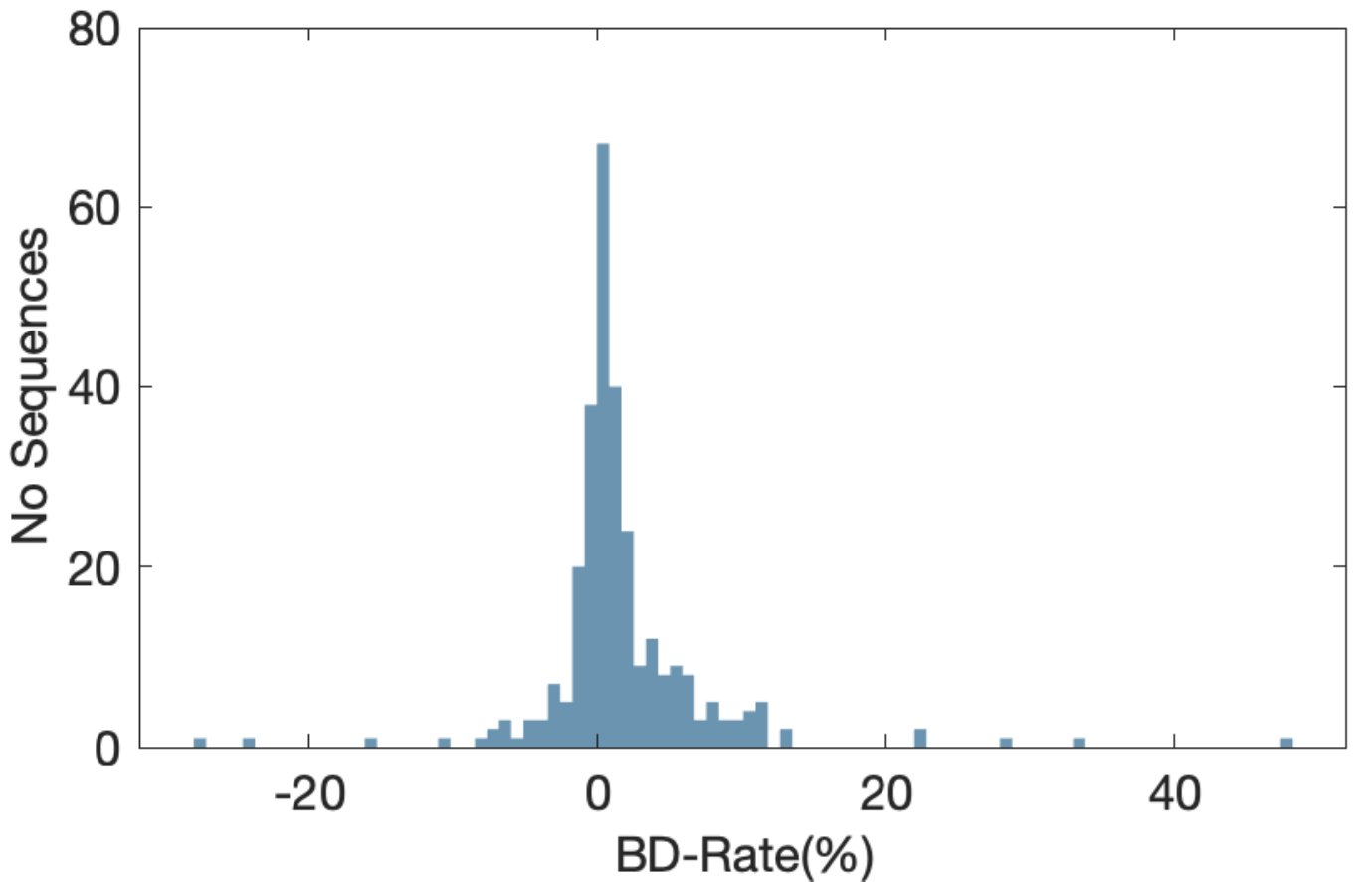}
\footnotesize{(c) FL vs RL method - BD-Rate: mean 1.6846\%, mad 3.2778\%.}
%\vspace{.1cm}
 \end{minipage}
 \hfill
 \begin{minipage}{.49\linewidth}
\centering    
\includegraphics[width=\linewidth]{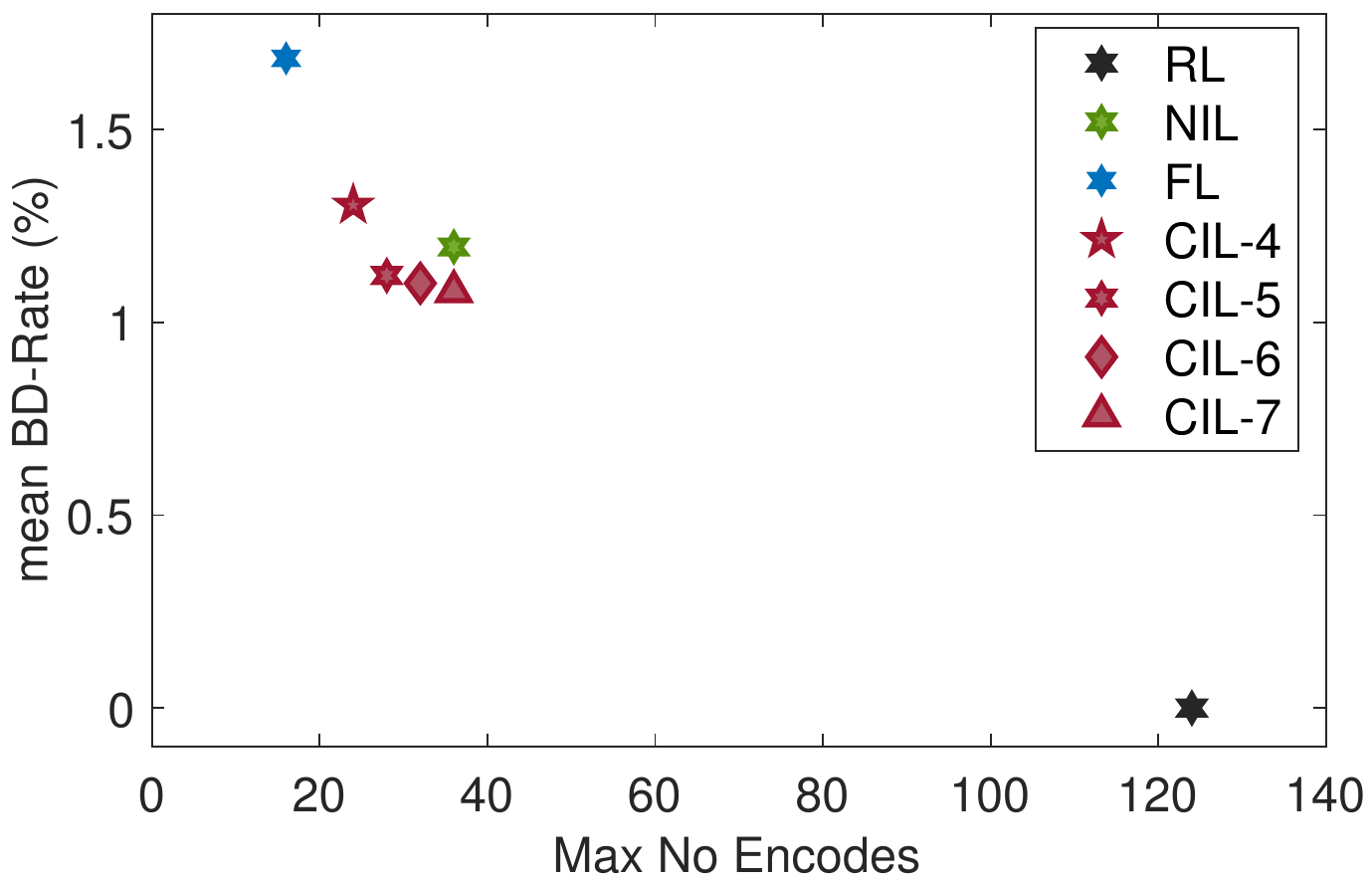}
\footnotesize{(d) mean BD-Rate vs maximum number of encodes.}
%\vspace{.1cm}
 \end{minipage}
\caption{The BD-Rate histograms for the compared methods and complexity.}
\vspace{-0.3cm}
\label{fig:BDRateHists}
\end{figure}

Fig.~\ref{fig:BDRateHists} illustrates the distributions of BD-Rates for the three tested methods in (a)-(c) and the complexity - accuracy tradeoff in (d). As can be seen in (d), for the same number of initial encodes, CIL-7, slightly improves the mean BD-Rate (0.13\%). As the differences between the CIL versions 5-7 are trivial, it is clear that CIL-5 results in the best accuracy-complexity tradeoff, as it reduces by 22\% the required encodes compared to NIL. Moreover, FL can further decrease the complexity down to only eight initial encodes at the cost of 0.46\% BD-Rate loss. However, by comparing the histograms, the FL distribution tail is heavier than the other two methods. 
Compared to RL, NIL reduces the maximum number of required encodes by 71\% with 75.1\% RL-hits, CIL-5 by 77.4\% with 74.3\% RL-hits, and FL by 87.1\% with 36\% RL-hits. Taking into account these statistics, we conclude that CIL-5 is the recommended method for an accurate and cost-effective ladder estimation.

\begin{figure}[!t]
\begin{minipage}[b]{.49\linewidth}
\centering
\includegraphics[width=\linewidth]{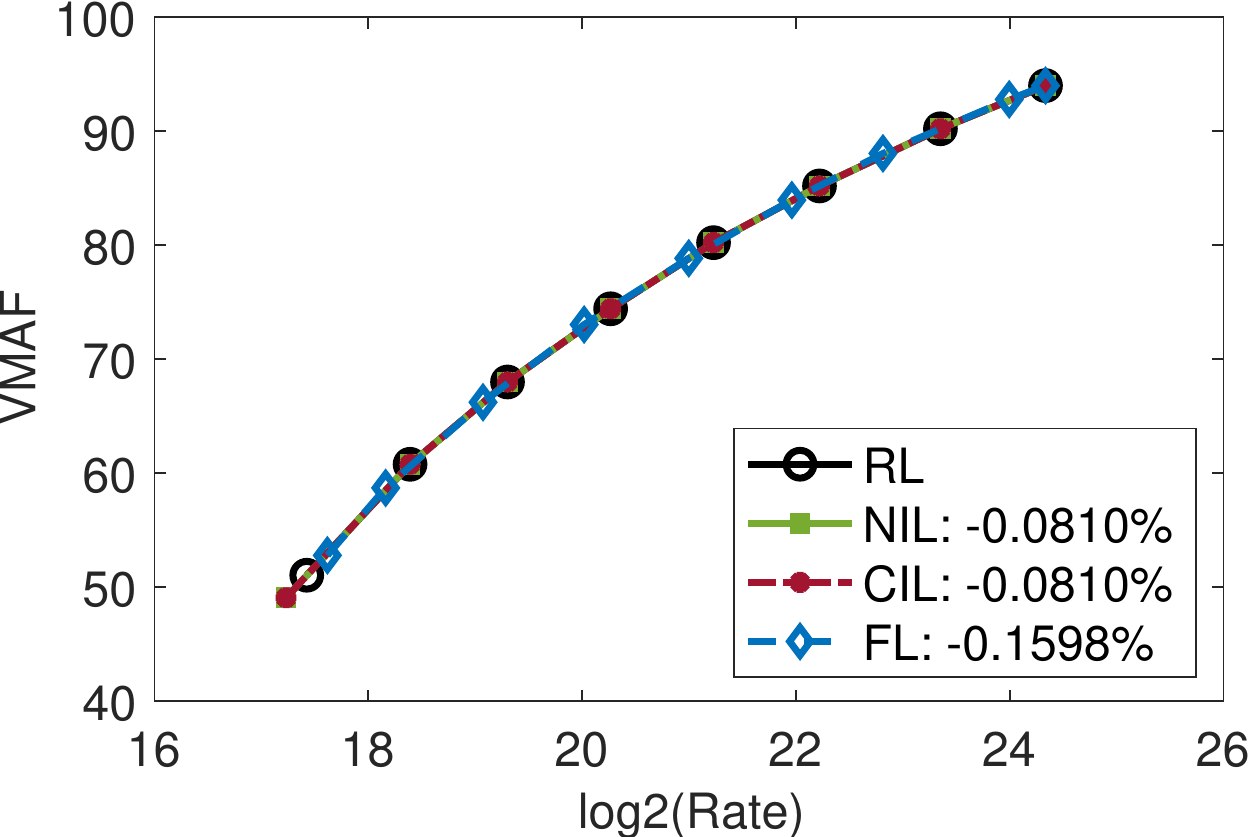}
\footnotesize{(a) Air-acrobatics.}
%\vspace{.1cm}
\end{minipage}
\hfill
\begin{minipage}[b]{.49\linewidth}
\centering
\includegraphics[width=\linewidth]{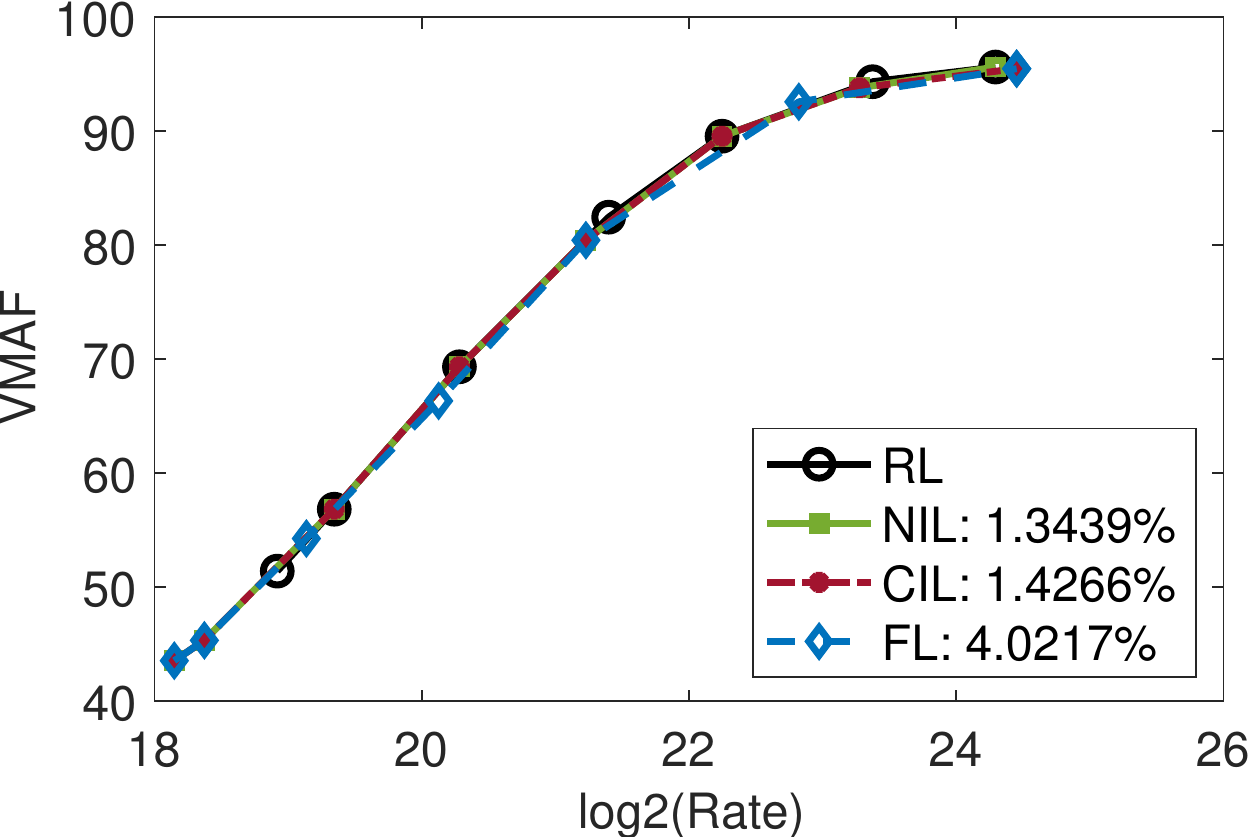}
\footnotesize{(b) ConstructionField.}
%\vspace{.1cm}
\end{minipage}

\begin{minipage}[b]{.49\linewidth}
\centering
\includegraphics[width=\linewidth]{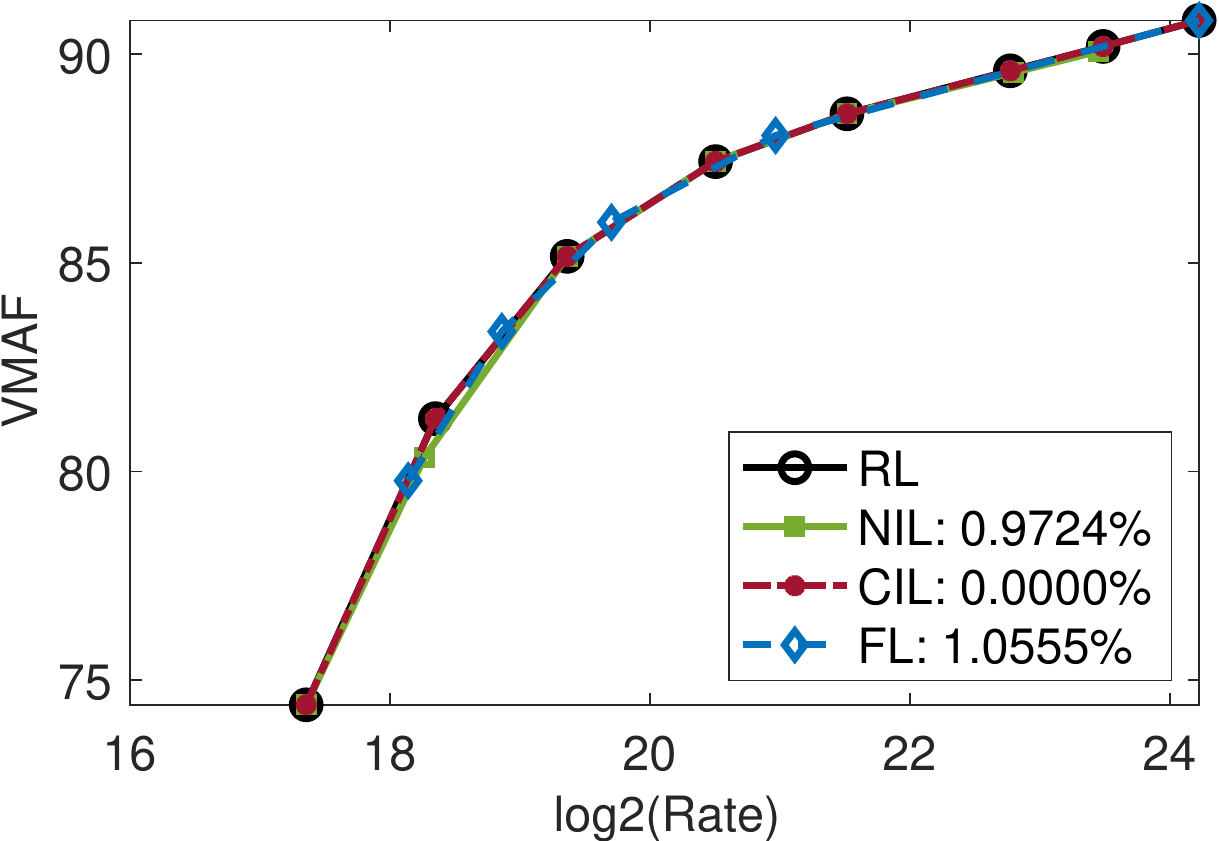}
\footnotesize{(c) Dinnerscene-scene2.}
%\vspace{.1cm}
\end{minipage}
\hfill
\begin{minipage}[b]{.49\linewidth}
\centering
\includegraphics[width=\linewidth]{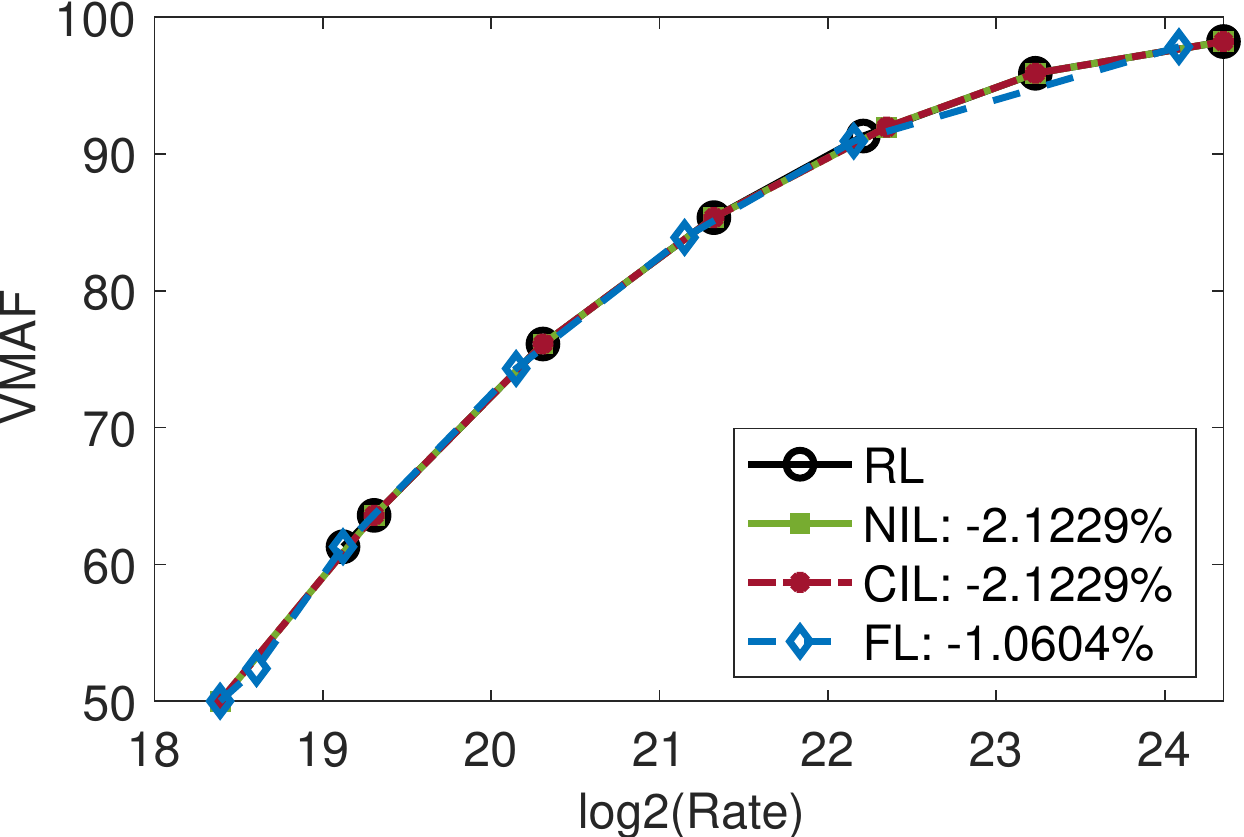}
\footnotesize{(d) WindAndNature.}
%\vspace{.1cm}
\end{minipage}
\caption{Predicted ladders for a selection of sequences that represent the distribution of BD-Rates in Fig.~\ref{fig:BDRateHists}.}
\vspace{-0.7cm}
\label{fig: ResultsSeqs}
\end{figure}

In Fig.~\ref{fig: ResultsSeqs}, we show a few examples of the resulting bitrate ladders. As can be seen, in most cases the performances of CIL and NIL are very similar. Although FL builds in most cases a bitrate ladder with PF points, these points are shifted, as also indicated by the RL-hits figure. On average, NIL and CIL are more successful in building ladders with points identical to RL points. 

\section{Conclusion}
\label{sec: Conclusion}
In this paper we proposed a content-driven method that can predict the bitrate ladder for adaptive streaming with significantly reduced complexity. CIL exploits spatio-temporal features extracted from uncompressed video to predict the Rate-VMAF curvature, in order to guide an interpolation based method towards the range of QP values that reside on the PF. 
The results showed a significant reduction of complexity, 77.4\% at a small BD-Rate cost, 1.12\%, when compared to the optimal reference ladder, while achieving to build a ladder with 74.3\% RL-hits.
Concluding, CIL with five initial encodes per resolution offers the best complexity-accuracy tradeoff and is therefore the recommended method for large-scale systems.

%\clearpage
%\newpage
\bibliographystyle{IEEEbib}
\bibliography{refs1}

\end{document}